\documentstyle[preprint,aps,psfig]{revtex}
\draft
\input{epsf}
\begin{document}
\title{
       \vspace*{-0.5cm}\hfill {\tt cond-mat/9507xxx}
       \vspace*{0.5cm}
       \\
Roughness of randomly crumpled elastic sheets}
\author{Frank Tzschichholz, Alex Hansen\footnote{Permanent address:
Institutt for fysikk, Norges tekniske h{\o}gskole, N--7034 Trondheim,
Norway} and St\'ephane Roux}
\address{Laboratoire de Physique et M\'ecanique des
Milieux H\'et\'erog\`enes,\\
Ecole Sup\'erieure de Physique et Chimie Industrielles,\\
10 rue Vauquelin, F--75231 Paris Cedex 05, France}
\date{\today}
\maketitle
\begin{abstract}
We study the roughness of randomly crumpled elastic sheets.  Based on
analytical and numerical calculations, we find that they are self affine
with a roughness exponent equal to one.  Such crumpling occurs {\it e.g.\/}
when wet paper dries.  We also discuss the case of correlated crumpling,
which occurs in connection with flattening of randomly folded paper.
\end{abstract}
Take a sheet of paper. Wet it with water and let it dry.  The sheet will
crumple.\footnote{The baking of traditional Norwegian flatbread or
Indian {\it papadam\/} produces a much stronger crumpling, but of the same
kind.} A closer look at the dried surface does not reveal any preferred
length scale in the crumpling.  Rather, smaller crumples appear inside the
larger ones, and so on to smaller and smaller scales, finally to reach the
scale of the fibers making up the paper.  Surfaces having this appearance
are in fact abundant in nature.  For example, fracture surfaces show
roughness on all scales, as do surfaces that have been corroded
\cite{fv91,m93,hz94,bs95}.
Recently, mathematical tools have been developed to describe such surfaces
in terms of their scaling properties \cite{m82,f88}. We will in this
letter analyse the roughness of dried paper surfaces in light of these
scaling properties based on a theoretical model originating from the theory
of linear elasticity of thin plates \cite{l44,f65}.  We also note the
connection between this problem and the problem of unfolding a randomly
wrapped paper \cite{pr95}:  Wrap a sheet of paper randomly into a tight ball.
Then unfold it and try to flatten it as much as possible.  It will stay
rough by much the same mechanism that makes dried paper rough.  However,
the foldings of the paper have introduced long-range correlations between the
local crumpling of the paper.  These long-range correlations may change
the roughness of the sheets.  Experimental measurements of this roughness
show, however, that it is within the experimental uncertainty equal to
the roughness due to uncorrelated crumpling.  We generalize the
one-dimensional model of reference \cite{pr95} to a two-dimensional
elastic sheet.  However, the roughness found in this model is larger than
the one found in the random crumpling case and in the experiments.

The scaling properties alluded to above, are more precisely described
through the concept of {\it self affinity.\/} We choose for a given surface
the $(x,y)$ plane to be the mean plane, and the $z$ axis to be the normal
direction.  The surface is self affine if it is (statistically) invariant
under the rescaling transformation
\begin{equation}
\label{100}
\cases{x&$ \to     \lambda  x   $\cr
       y&$ \to     \lambda  y   $\cr
       z&$ \to \mu(\lambda) z\;,$\cr}
\end{equation}
where $\lambda$  is a arbitrary scaling factor, and $\mu$ is a
function of $\lambda$.  Group properties on the transformation (\ref{100})
imply that $\mu$ is a homogeneous function of $\lambda$ characterized by an
exponent $H$, $\mu \propto\lambda^H$. This exponent is the
{\it roughness index.\/}  The aim of this letter is to determine the
roughness index of a randomly crumpled elastic sheet.

We return to the problem of wetting and then drying paper. By pouring water
on one side of a sheet of paper, it immediately curls up away from the
wetted side.  This is due to the paper expanding locally on the wetted side.
If, however, the paper is wetted on both sides, a uniform expansion occurs,
and the now thoroughly wet paper will stay flat. However, as it begins to
dry, it roughens.  There are two mechanisms behind this roughening: (1)
internal torques which bend the surface locally due to inhomogeneous
contractions in the perpendicular direction of the sheet, which in turn
are due to inhomogeneities in the structure of the paper itself and/or
differences in the drying drying rates from the two sides of the paper, and
(2) buckling due to locally homogenous in-plane contractions of the paper.
In deciding which of these two mechanisms is the dominant one, we note that
buckling, which is a non-linear elastic problem, produce multiple stable
bending modes, while the first process to a first approximation is a linear
problem producing unique bending modes.  Using whether a unique roughened
shape of the paper results when the paper has dried, or there are several
equivalent ones, we find that mostly the former is the case.  We thus
conclude that it is the internal torques which dominate.

When wrapped paper is flattened, the resulting roughness is caused by
internal torques and not buckling.

In light of the above discussion, we base our modeling on the
linearized equation for flexion of an elastic plate under a loading
$q(x,y)$ in the $z$ direction:
\begin{equation}
\label{200}
D\Delta\Delta \zeta = q\;,
\end{equation}
where $D$ is the flexural rigidity of the plate, (chosen to be unity, $D=1$,
in the following), and $\zeta(\vec x )$ is the vertical deflection of the
plate at $\vec x = (x,y)$ \cite{f65}. Equation (\ref{200}) is known as the
Germain equation.\footnote{An interesting note on the history of this
equation and Sophie Germain can be found in \cite{t53}.}  However, the
forces acting on the surface in the case of drying paper are not external.
We are rather seeking a behavior where locally the surface at $\vec x$ would
adopt a non-zero curvature $\Delta \zeta=\eta(\vec x)$ if it were isolated
from its neighborhood. The $\eta(\vec x)$ is chosen here to be a random field.
This noise term is assumed to represent the fluctuations in curvature due to
the local inhomgeneous contractions of the paper.  However, $\Delta\zeta$
is not a force, $\Delta\Delta\zeta$ is, see equation (\ref{200}).  Therefore,
the equation that balances the forces locally in the plane is
\begin{equation}
\label{300}
\Delta\Delta \zeta=\Delta\eta\;.
\end{equation}
With the assumption that $\eta(\vec x)$ is an {\it uncorrelated\/} random
field, we have that $\langle\eta (\vec x_1)\eta(\vec x_2)\rangle=
c^2\delta(\vec x_1-\vec x_2)$, where $\langle \dots\rangle$ signify ensemble
average, and $\delta(\vec x)$ is the Dirac distribution.  Note, however,
that modeling the unfolded and flattened paper, the stochastic field $\eta$
has long range correlations build into it stemming from the folding process.

Let us assume a system of size $L\times L$ where $L$ is large.  The height
fluctuations, $\langle (\zeta(\vec x_1) -\zeta(\vec x_2))^2\rangle$, may
easily be calculated by using Green's method.  We have that
\begin{equation}
\label{400}
\zeta(\vec x)=\int d\vec y\> G(\vec x - \vec y)\Delta_{y}\eta(\vec y)
=\int d\vec y\>\Delta_{x} G(\vec x - \vec y)\eta(\vec y)\;,
\end{equation}
where $G(\vec x)$ is the solution of $\Delta\Delta \zeta
=\delta(\vec x)$, which in Fourier space is $\tilde G(\vec k) =1/k^4$. Thus,
we have that
\begin{eqnarray}
\label{500}
\langle \zeta(\vec x_1) \zeta(\vec x_2)\rangle&=&\int\int
d\vec y_1 d\vec y_2\>
\Delta_{x_1} G(\vec x_1 - \vec y_1)
\Delta_{x_2} G(\vec x_2 - \vec y_2)
\langle\eta(\vec y_1)\eta(\vec y_2)\rangle\\
&=&c^2 \int k^4 \> d\vec k\>\tilde G(\vec k)\tilde G(-\vec k)
e^{i (\vec x_1-\vec x_2)\cdot \vec k}
=c^2 \int {{d\vec k}\over {k^4}} e^{i (\vec x_1-\vec x_2)\cdot \vec k}
\;.\nonumber
\end{eqnarray}
We may now calculate the height fluctuations:
\begin{equation}
\label{550}
\langle (\zeta(\vec x_1)-\zeta(\vec x_2))^2\rangle=
2c^2 \int_0^{2\pi}d\phi\>\int_{1/L}^\infty {{dk }\over {k^3}}\>
\left(1-e^{i \vert\vec x_1-\vec x_2\vert k\cos(\phi)}\right)
=C L |\vec x_1-\vec x_2|^2\;,
\end{equation}
where $C$ is a constant.  This last result shows that the roughness
index is
\begin{equation}
\label{700}
H=1\;.
\end{equation}
One should also note that the variance of the height scales as
\begin{equation}
\label{710}
\langle\zeta^2(\vec x)\rangle \propto c^2L^2
\end{equation}
with the system size. The square dependence on $L$ is also a reflection of
$H=1$.

A direct way to model numerically surfaces following equation (\ref{300}),
is to integrate this equation to get
\begin{equation}
\label{725}
\Delta\zeta=\eta+\phi\;,
\end{equation}
where $\phi$ is a solution of the equation $\Delta \phi=0$.  We now assume
that our system of size $L\times L$ has biperiodic boundary conditions.  Then,
we have that $\langle\Delta\zeta\rangle=0$ if the field $\zeta$ is to be
continuous.  Equation (\ref{725}) then leads to the condition
$\langle\phi\rangle=-\langle\eta\rangle$ on $\phi$.  The solution of the
Laplace equation with biperiodic boundary conditions is a constant, $\phi=
constant=-\langle\eta\rangle$.  Thus, we have in Fourier space that
\begin{equation}
\label{750}
\tilde\zeta={1\over k^2}\left(\tilde\eta-\langle\eta\rangle\right)\;.
\end{equation}
This is straight forward to implement on the computer.

In addition to generating the solution of equation (\ref{300}) through
equation (\ref{750}), we have discretized (\ref{300}) and solved it through
an iterative method.  This method raises principal questions in that the
noise term $\eta$ is not a smooth function, but varies rapidly on all
scales, including the one chosen as our lattice constant in the
discretization.  This is a typical problem faced when modeling strongly
disordered systems.  However, we note that {\it a priori\/} a continuum
description of a strongly disordered system (where this is at all possible)
is not more fundamental than a description based on a discrete system.

\begin{figure}
\centerline{
        \epsfxsize=14.0cm
        \epsfbox{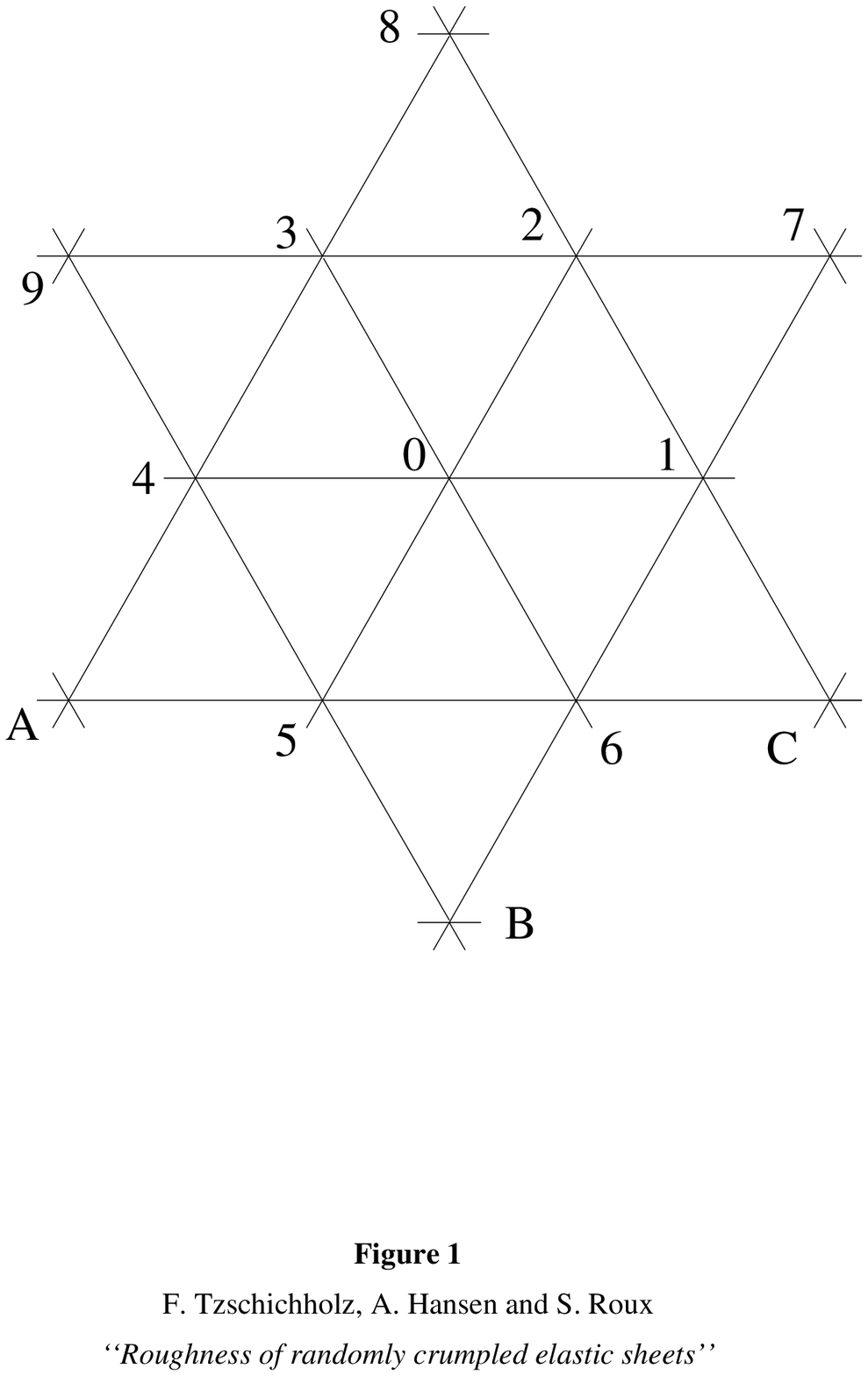}
        \vspace*{0.5cm}
        }
\caption{\label{fig1} The numerical calculation is done using a triangular
tiling of rhombical elastic plaquettes as shown in this figure.}
\end{figure}

The discretization we have chosen is the following one:  We use a triangular
tiling of rhombical plaquettes as shown in figure \ref{fig1} \cite{rh88}.
The flexural energy associated with plaquette $0394$ (the numbering refers
to the nodes of figure \ref{fig1}) is
$\epsilon_{0394}=(\zeta_0-\zeta_3+\zeta_9-\zeta_4+\eta_{34})^2$, where
$\eta_{34}$ corresponds to the noise $\eta$ of equation (\ref{300}).  Each
elementary triangle is covered by three plaquettes, {\it e.g.\/} triangle
$034$ is the intersection of plaquettes $0394$, $0345$ and $0234$.  With
this construction, the dependence of the elastic energy of the network on a
given node 0 in figure \ref{fig1} is given by the sum
\begin{eqnarray}
\label{1000}
\epsilon_0&=&(\zeta_0-\zeta_1+\zeta_7-\zeta_2-\eta_{12})^2
            +(\zeta_0-\zeta_2+\zeta_8-\zeta_3-\eta_{23})^2\nonumber\\
          &+&(\zeta_0-\zeta_3+\zeta_9-\zeta_4-\eta_{34})^2
            +(\zeta_0-\zeta_4+\zeta_A-\zeta_5-\eta_{45})^2\nonumber\\
          &+&(\zeta_0-\zeta_5+\zeta_B-\zeta_6-\eta_{56})^2
            +(\zeta_0-\zeta_6+\zeta_C-\zeta_1-\eta_{61})^2\\
          &+&(\zeta_0-\zeta_1+\zeta_2-\zeta_3+\eta_{02})^2
            +(\zeta_0-\zeta_2+\zeta_3-\zeta_4+\eta_{03})^2\nonumber\\
          &+&(\zeta_0-\zeta_3+\zeta_4-\zeta_5+\eta_{04})^2
            +(\zeta_0-\zeta_4+\zeta_5-\zeta_6+\eta_{05})^2\nonumber\\
          &+&(\zeta_0-\zeta_5+\zeta_6-\zeta_1+\eta_{06})^2
            +(\zeta_0-\zeta_6+\zeta_1-\zeta_2+\eta_{01})^2\;.\nonumber
\end{eqnarray}
The total elastic energy of the network is the sum
$E=\sum_{i=1}^N\epsilon_i/12$, where $i$ runs over all $N$ nodes.  In the
limit of a vanishing lattice constant, $\delta$,
$\partial\epsilon_0/\partial\zeta_0=0$ leads to the equation
\begin{equation}
\label{1050}
\delta^4\Delta\Delta\zeta={{\delta^2}\over{12}}\Delta\eta
+{{4\delta^4}\over 3}\Delta\Delta\eta+\dots\;.
\end{equation}
When $\eta$ is a smooth function in the scale of $\delta$, the second term
on the right hand side of equation (\ref{1050}) is small compared to the
first term.  However, if $\eta$ is not smooth, it will be of the same order
as $\Delta\sim1/\delta^2$, and the second term (and those we have not
explicitly written in (\ref{1050})) will be of the same size as the first.
This is the case we deal with here.  However, even though the implementation
(\ref{1000}) does not reproduce (\ref{300}) on small scale when $\eta$ is
not smooth, the large scale features will be the same as predicted by the
continuum equations. Had our basic equations
not been linear, however, such a conclusion could not have been drawn. We
note, furthermore, that equation (\ref{1000}) is an equally accurate but
distinct modelization of the original problem, as the one based on the
continuum theory, (\ref{300}).

\begin{figure}
\centerline{
        \epsfxsize=14.0cm
        \epsfbox{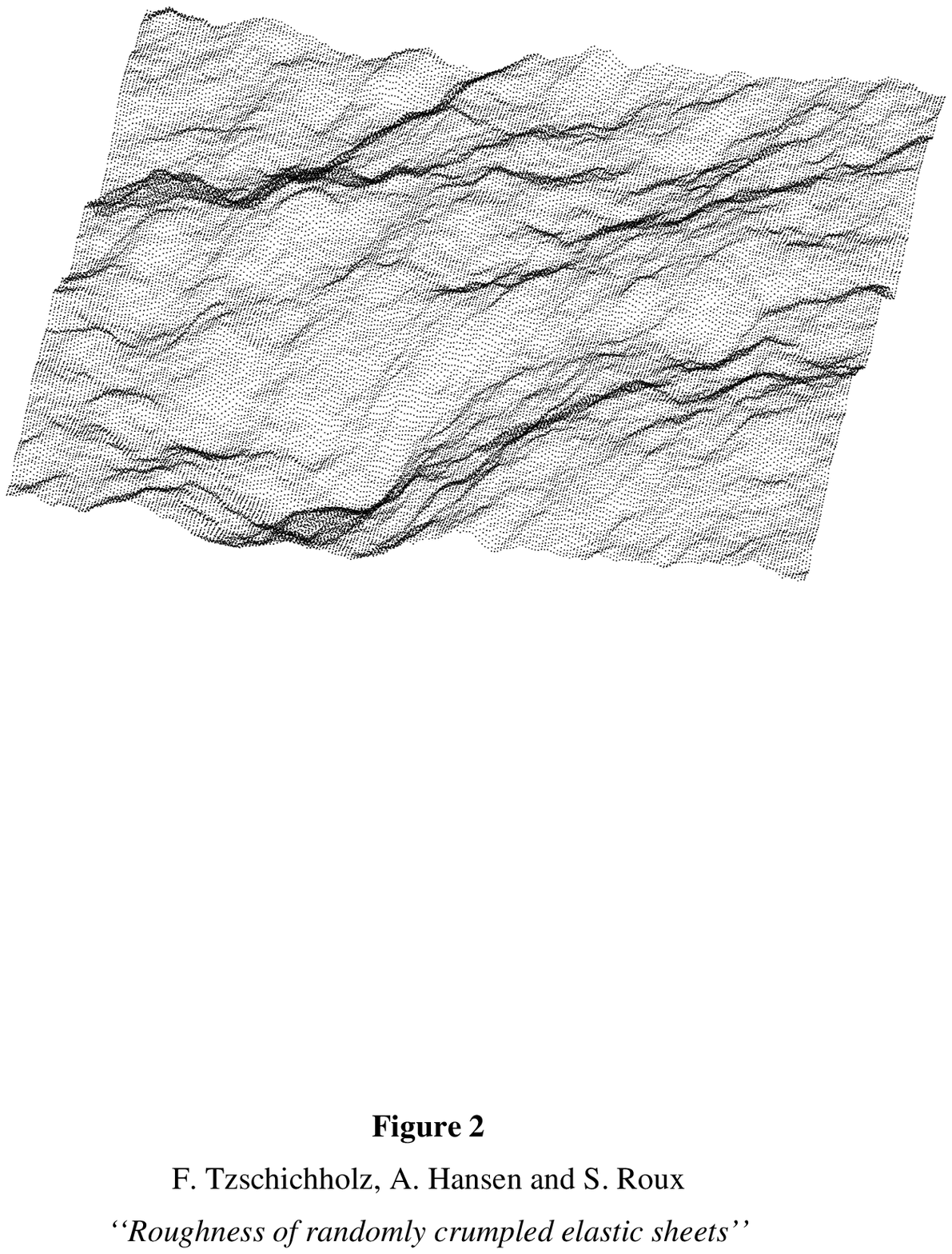}
        \vspace*{0.5cm}
        }
\caption{\label{fig2} A randomly crumpled sheet of size $256\times 256$.}
\end{figure}

The noise term $\eta_{ij}$ in (\ref{1000}) is associated with the bond $ij$,
which in turn is associated with a plaquette.  We choose the value of
$\eta_{ij}$ to be either 1 or $-1$ with equal probability.  We implement
biperiodic  boundary conditions.  For a given distribution of noise terms
$\eta_{ij}$, we minimize the total energy $E$ using a conjugate gradient
algorithm as described in \cite{bh88}.  In figure \ref{fig2}, we show a
typical crumpled surface of size $L\times L=256\times 256$.

\begin{figure}
\centerline{
        \epsfxsize=14.0cm
        \epsfbox{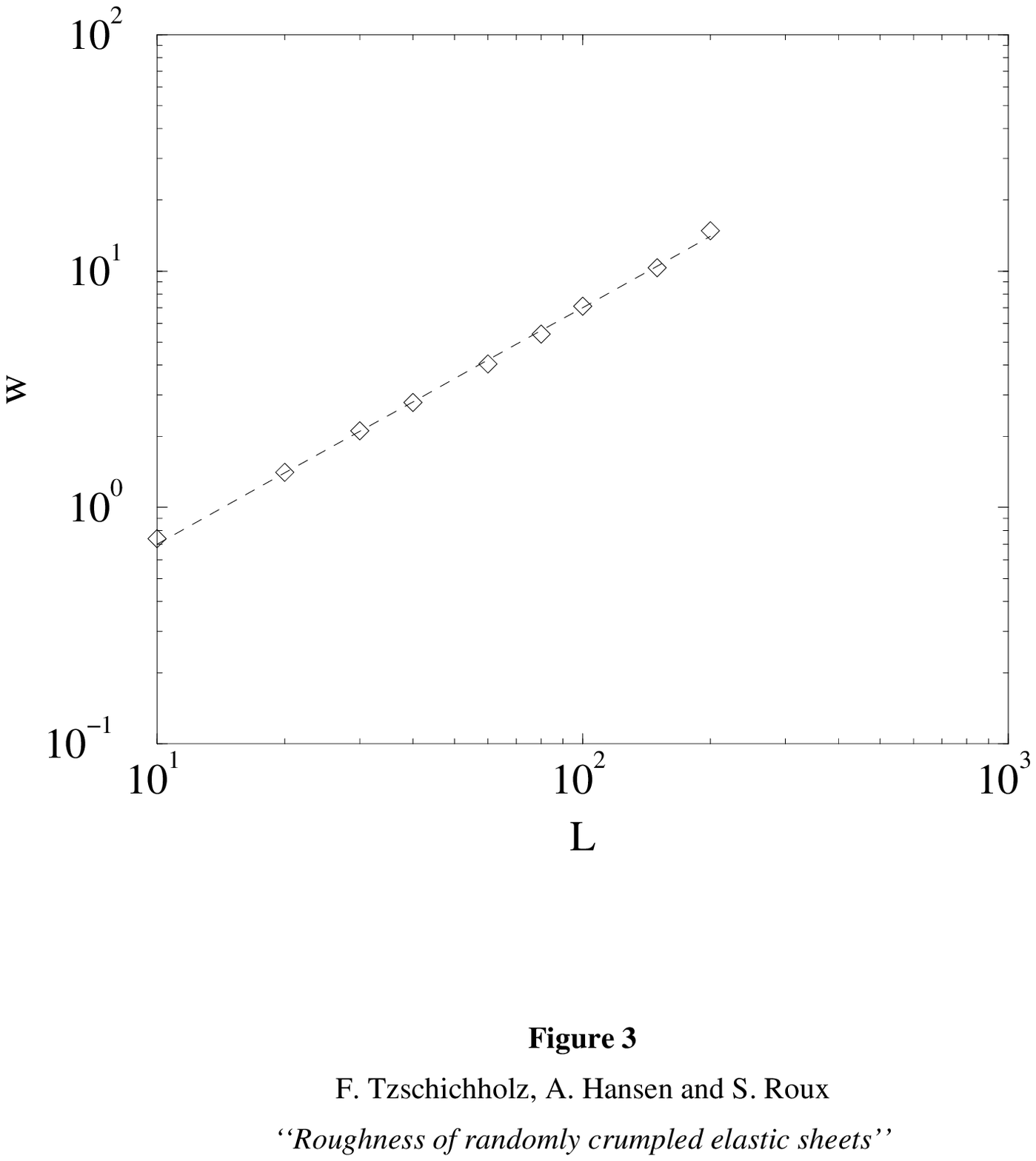}
        \vspace*{0.5cm}
        }
\caption{\label{fig3} Average width $w$ of sheets of size $L\times L$. The
slope of the broken line is 1.0.}
\end{figure}

\begin{figure}
\centerline{
        \epsfxsize=14.0cm
        \epsfbox{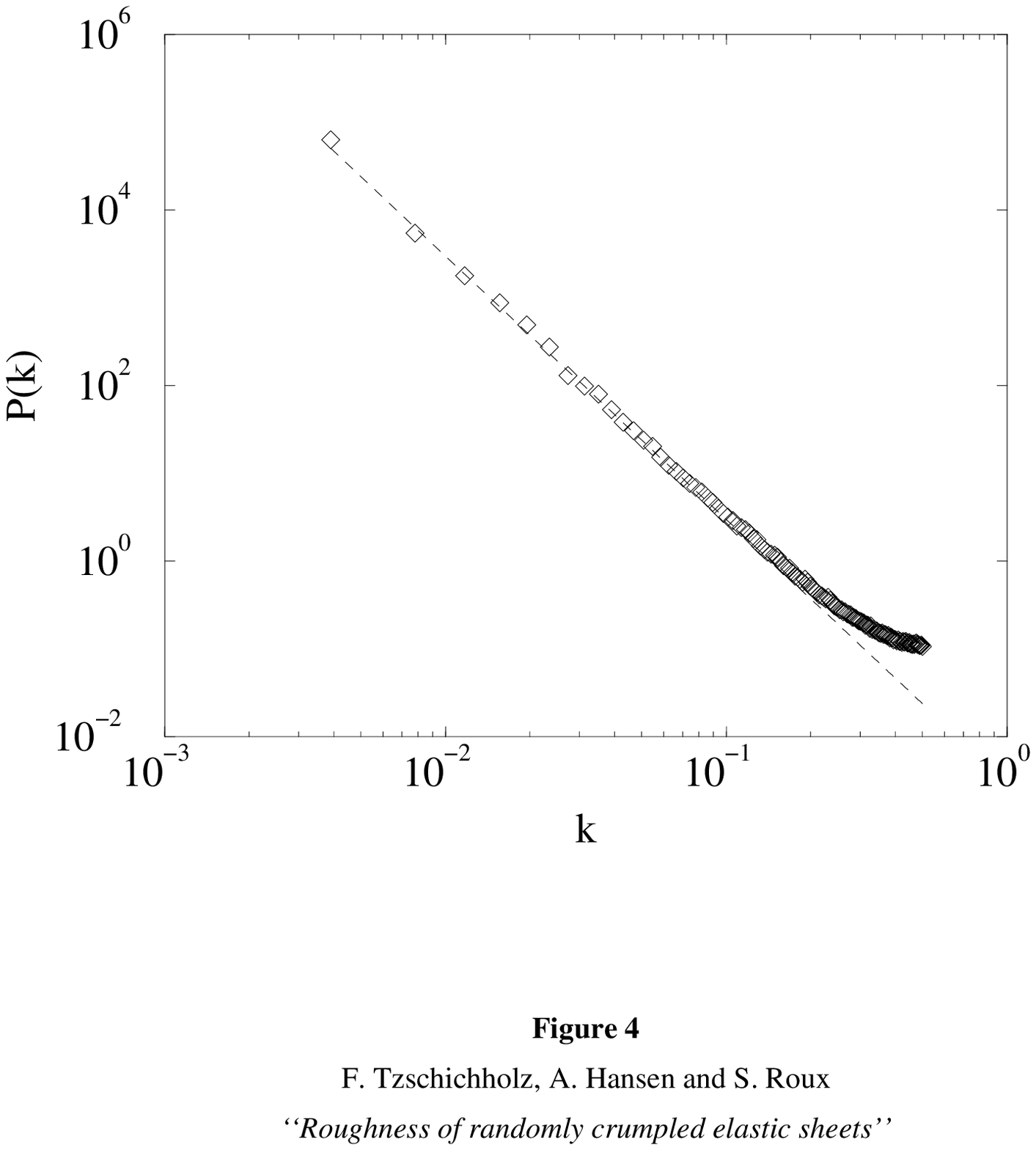}
        \vspace*{0.5cm}
        }
\caption{\label{fig4} Power spectrum $P(k)$ as a function of wave number
$k$ measured along one-dimensional profiles of surfaces that have been
randomly crumpled. The curve is an average over all 256 profiles of each
of 10 samples of size  $256\times 256$. The slope of the broken line is
$-3.0$.}
\end{figure}

In order to measure the roughness index we generated an ensemble
consisting of 1000 networks of size $L=10$, 1000 samples of size $L=20$,
800 samples of size $L=30$, 600 samples of size $L=40$, 200 samples of
size $60$, 80 samples of size $L=80$, 40 samples of size $L=100$, 20
samples of size $L=150$, and 10 samples of size $L=200$.  For each sample we
measured the width defined as $w^2=(\sum_i \zeta_i^2)/N-
(\sum_i\zeta_i)^2/N^2$, which corresponds to (\ref{710}), and then averaged
this quantity over all samples
with the same size $L$.  The result is shown in figure \ref{fig3}.  We find
that $w\propto L^H$ with $H=1.0\pm0.05$, in agreement with the predicted
$H=1$.  In figure \ref{fig4} we show the power spectrum $P(k)$, which is the
Fourier transform of the height-height correlation function, measured along
one-dimensional profiles and averaged over 10 samples of size
$256\times 256$.  $k$ is the wave number. For a self-affine surface,
one expects $P(k)\propto k^{-(1+2H)}$.  We find a slope equal to
$-3.0\pm 0.1$ in figure \ref{fig4}, again consistent with a roughness index
equal to one.

Thus, we have succeded in demonstrating how a simple linear elastic theory
of crumpled surfaces can generate non-trivial self-affine surfaces.  It
would be very interesting to see how these theoretical results match with
experiments.  We furthermore note that we may use these results to generate
large self-affine surfaces by {\it e.g.\/} spotwise random heating of
elastic plates.

\begin{figure}
\centerline{
        \epsfxsize=14.0cm
        \epsfbox{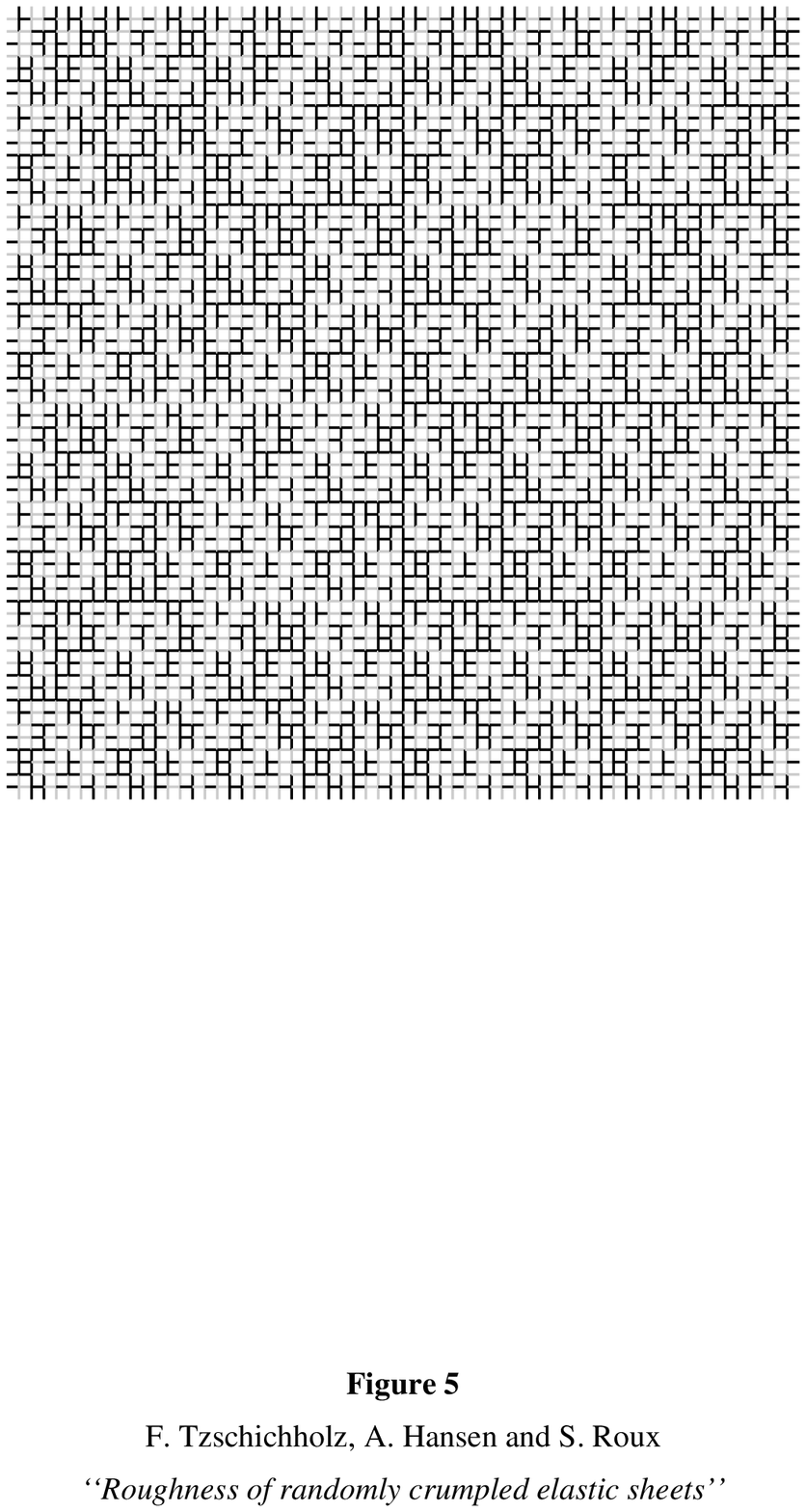}
        \vspace*{0.5cm}
        }
\caption{\label{fig5} A folding pattern generated by the algorithm described
in the text.  Black line means the bend is ``upwards" while a gray line means
that the bend is ``downwards." The size here is $64\times 64$.}
\end{figure}

We now turn to the problem of randomly folded paper.  The folding process
induces in the paper a bending along the fold line somewhat similar to
that observed when wet paper dries.  Suppose now we fold the paper a first
time.  This produces a straight bend.  However, the next fold --- when the
paper is unwrapped again --- will change direction when it meets the first
fold, unless this second folding is perpendicular to the first one.  In
addition, the bending of the second fold will change ``sign" when it
crosses the first fold.  Thus, taking as starting point the one dimensional
geometrical model introduced in reference \cite{pr95}, we implement the
following two-dimensional model.  We associate with each plaquette in a
square lattice of size $2^N\times 2^N$ a value $\eta$.
Assign to the line of plaquettes with coordinates $(2^(N-1),j)$, where
$j=1$, ... $2^N$, equal values for $\eta$, either $+1$ or $-1$.  Then,
go to line $(i,2^{N-1})$, where $i=1$, ..., $2^{N-1}$.  Assign to the
first node an $\eta$-value equal to either $+1$ or $-1$ with equal
probability.  This value is reproduced for all subsequent nodes along this
line until the previous line is encountered at $(i=2^{N-1},2^{N-1})$. Then,
for all $(i>2^{N-1},2^{N-1})$, we assign the opposite value for $\eta$.  We
then go to the line $(2^{N-2},j)$, choose an $\eta$-value equal to $\pm 1$
for node $(2^{N-2},1)$, reproduce this value until we cross an older line,
and then change sign for the subsequent $\eta$ values.  Each time an older
line is met, we change sign in the assigning of $\eta$ values.  This
construction is repeated over and over until all nodes has a value for $\eta$
associated with it.  We show in figure \ref{fig5} a pattern of bending lines
produced this way for a $64\times 64$ lattice.  Black line means an
associated $\eta=1$, and gray means that the associated $\eta=-1$.

\begin{figure}
\centerline{
        \epsfxsize=14.0cm
        \epsfbox{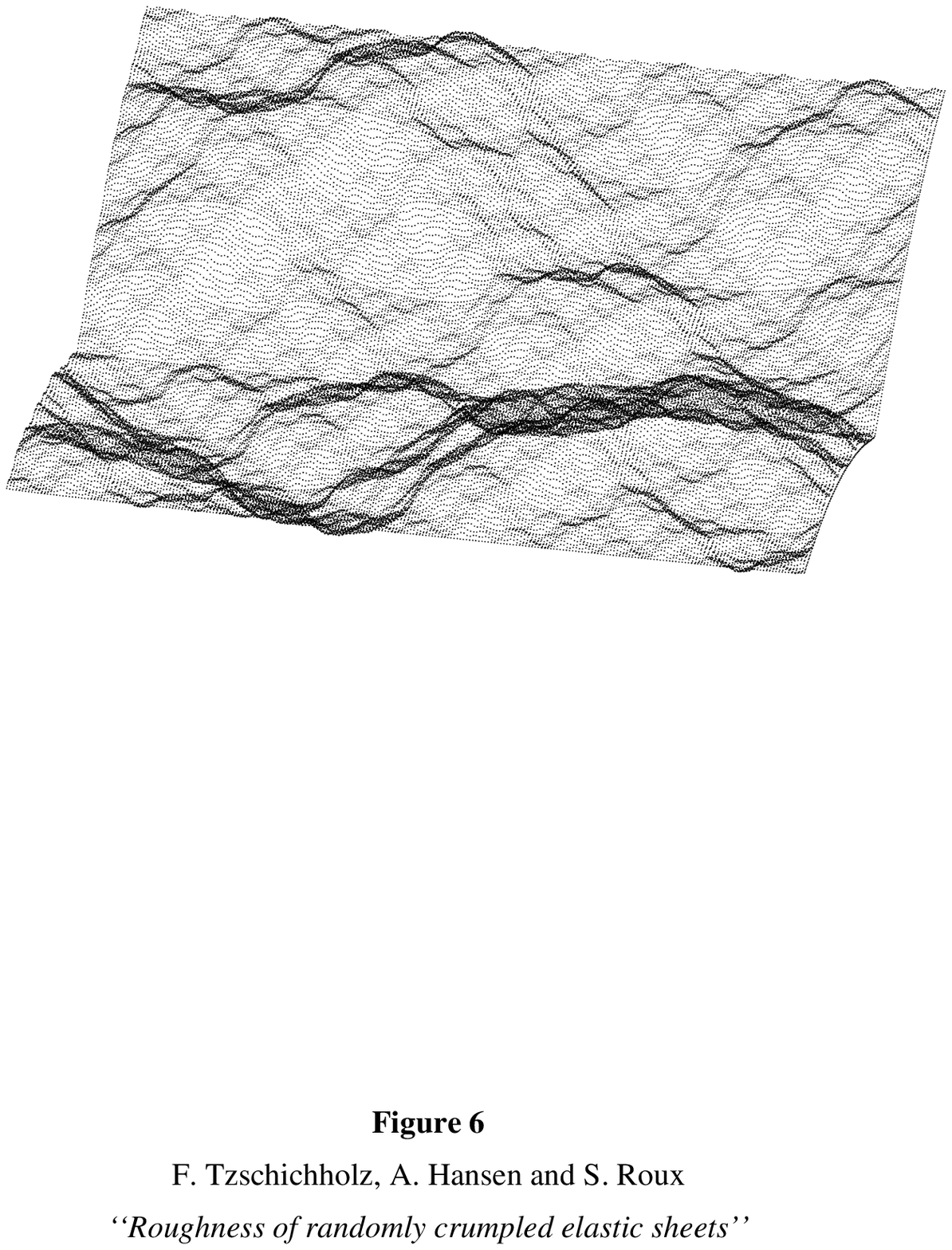}
        \vspace*{0.5cm}
        }
\caption{\label{fig6} A sheet that has been been crumpled through the
random folding algorithm.  The size is $256\times 256$.}
\end{figure}

\begin{figure}
\centerline{
        \epsfxsize=14.0cm
        \epsfbox{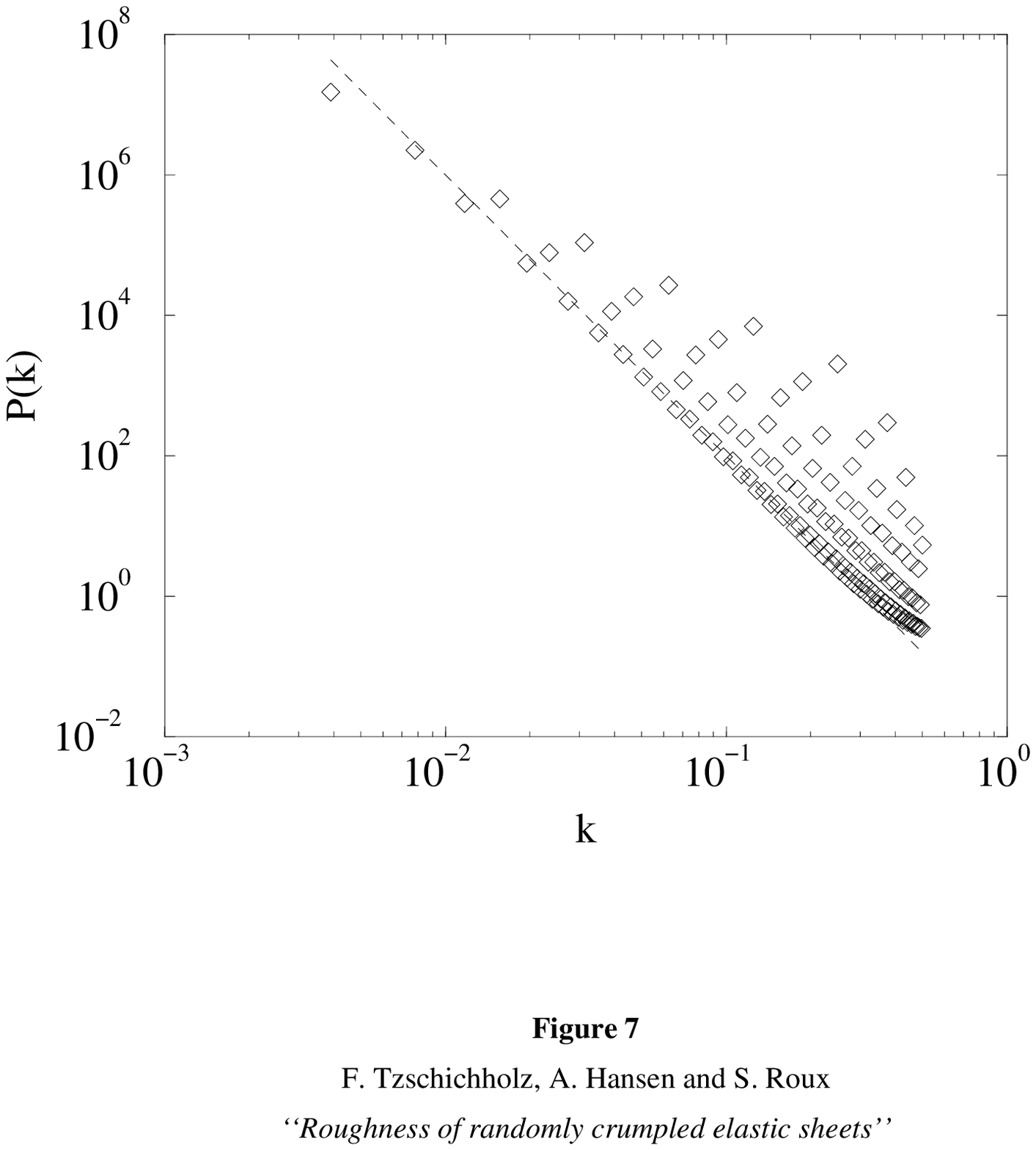}
        \vspace*{0.5cm}
        }
\caption{\label{fig7} Power spectrum $P(k)$ as a function of wave number
$k$ for the randomly folded surfaces.  We measured along one-dimensional
profiles over 20 samples of size $256\times 256$.
The slope of the broken line is $-4.0$.}
\end{figure}

We then proceed to solve equation (\ref{300}) with this distribution of
$\eta$ values.  In practice, we used the solution given in equation
(\ref{750}). We show in figure \ref{fig6} a $256\times 256$ sheet
produced this way.  We note the similarity with figure \ref{fig2} that
was generated from a random distribution of $\eta$ values.  However,
an analysis of the power spectrum of an ensemble of such lattices (20
samples of size $256\times 256$) gave the result shown in figure \ref{fig7},
where we plot on log-log scale $P(k)$ {\it vs.\/} $k$ in the same way
as in figure \ref{fig4}.  There is a rich structure in this power spectrum
which reflects the correlations induced by the folding process.  Attempting
to extract a power law from this structure, we show astraight line that
follows the densest region of the plot.  This curve has a slope of $-4$,
thus indicating a roughness index equal to $1.5\pm0.1$ (from $1+2H=4$).
This is appreciably larger than what was found for the randomly crumpled
paper, $H=1$, and the experimental study of the randomly folded paper
\cite{pr95}: $Hi\approx 0.9$.

We have, thus, shown in this letter how a random internal bending process
of an elastic plate leads to a self affine structure with roughness index
equal to one.  Such processes are met {\it e.g.\/} when wet paper dries.
We have also studied the case of a correlated bending process,  {\it i.e.\/}
when paper is crumpled into a tight ball and then unfolded and flattened.
We find through a simple model for the correlations a roughness index which
is higher than the one found in the random crumpling case.

F.T.\ thanks Lafarge-Copp\'ee and the CNRS through the {\it Groupement de
Recherche Milieux H\'et\'erog\`enes Complexes\/} for financial support.
A.H.\ thanks the ESPCI for a visiting professorship during which tenure this
work was done.  Further support came from the CNRS and the NFR through a
{\it Programme Internationale de Coop\'eration Scientifique.\/}

 \end{document}